THE SCHWARZSCHILD GEOMETRY AND THE BLACK HOLES

M. Cattani
Instituto de Fisica, Universidade de S. Paulo, C.P. 66318, CEP 05315−970
S. Paulo, S.P. Brazil. E−mail: mcattani@if.usp.br

Abstract.

In this article we analyze the predictions of the Einstein gravitation theory (EGT) on black holes in the framework of the Schwarzschild geometry that is defined in the vacuum around a spherically symmetric mass distribution, without charge, not in rotation. The Eddington and Kruskal metrics have been also taken unto account and the topological connections named wormholes have been analyzed. This article was written to graduate and postgraduate students of Physics.
Key words: Einstein gravitation theory; Schwarzschild black holes;wormholes.

Resumo.

Nesse artigo analisamos as previsões da teoria de gravitação de Einstein (TGE) sobre os buracos negros no contexto da geometria de Schwarzschild que é definida no vácuo em torno de uma distribuição esfericamente simétrica de massa, sem carga, não em rotação. As métricas de Eddington e de Kruskal também foram levadas em conta e as conexões topológicas denominadas de "buracos de minhocas" foram analisadas. Esse artigo foi escrito para alunos de graduação e pós−graduação de Física.

## I.Introdução

Seguindo o procedimento adotado em nossos artigos anteriores[1] procuraremos citar um mínimo possível de referências (artigos e livros) fazendo os cálculos com suficiente precisão, deixando de lado alguns refinamentos que poderão ser encontrados em artigos. As previsões serão comparadas com resultados experimentais, quando esses existirem, sem a preocupação exagerada de analisar em detalhes as técnicas empregadas e suas limitações. Esses aspectos poderão ser vistos nas referências citadas.

Em um artigo precedente[1] usando a Teoria de Gravitação de Einstein (TGE) mostramos como calcular a métrica do espaço−tempo gerada no vácuo em torno de uma distribuição esfericamente simétrica de massa M, sem carga, em repouso e não em rotação. Essa métrica define a geometria de Schwarschild. Em coordenadas polares $x_o = x_4 = ct$, $x_1 = r$, $x_2 = \theta$ e $x_3 = \varphi$ essa métrica que é definida através do invariante $ds^2 = c^2 d\tau^2$ ($\tau$ = tempo próprio), dado por

$$ds^2 = (1- 2\kappa/r)c^2dt^2 - dr^2/(1 -2\kappa/r) - r^2d\theta^2 - r^2\sin^2\theta\, d\varphi^2 \qquad (I.1),$$

onde $\kappa = GM/c^2$, é chamada de métrica de Schwarschild (MS). Assim, de acordo com (I.1) temos $g_{oo}(r) = g_{44}(r) = Z = (1 - 2\kappa/r) = e^{N(r)}$, $g_{11}(r) = 1/Z = e^{M(r)}$ , $g_{22}(r) = r^2$ e $g_{33}(4) = r^2\sin^2\theta$.

A quantidade adimensional $\kappa/r=GM/c^2r$ pode ser vista como uma medida da intensidade do campo gravitacional[2]. Estudamos em artigos anteriores[3] vários efeitos gravitacionais quando $GM/c^2r << 1$ tais como a deflexão da luz, o efeito Doppler luminoso, a delação temporal, a precessão do periélio dos planetas e o retardo temporal de ecos de radar[3]. No sistema solar os efeitos gravitacionais relativísticos são muito pequenos. Basta notar que na superfície do Sol $GM_S/c^2R_s \sim 10^{-6}$ pois $G/c^2 = 7.414\ 10^{-28}$ m/kg, a massa $M_S$ e o raio $R_s$ do Sol são iguais a $M_S = 2.3\ 10^{30}$ kg e $R_s = 6.96\ 10^5$ km, respectivamente. Esses efeitos começam a ser tornar grandes nas vizinhanças de uma estrela muito massiva e muito compacta quando $2GM/c^2r$ se aproxima da unidade.[4] Por exemplo, quando $r = 3GM/c^2$ a deflexão da luz começa a se tornar tão grande que o sinal luminoso poderá de mover em uma órbita circular fechada[2] ao redor da estrela (Apêndice A).

J.Mitchell[5] em 1784 foi o primeiro a relatar o efeito espetacular produzido pelo potencial gravitacional GM/r quando ele se torna muito grande. Usando a teoria de gravitação de Newton (TGN) ele mostrou que a velocidade de escape $V_e$ de um corpo de massa m deve ser $V_e \geq (2GM/R)^{1/2}$ onde R é o raio do planeta. Como $V_e$ independe da massa do corpo ele argumentou que nem mesmo a luz poderia escapar da atração gravitacional se R fosse menor do que valor limite $r_s$ dado por

$$r_s = 2GM/c^2 \qquad (I.2).$$

Em termos de $r_s$ dada por (I.2) a MS (I.1) pode ser re−escrita na forma

$$ds^2 = (1- r_s/r)c^2dt^2 - dr^2/(1 - r_s/r) - r^2d\theta^2 - r^2\sin^2\theta\, d\varphi^2 \qquad (I.3).$$

Pode−se mostrar (Apêndice B) usando−se (I.3) que a curvatura do espaço−tempo $R^\alpha{}_{\beta\mu\nu}$ apresenta divergências no limite de $r \to r_s$. Como a componente $R^o{}_{1o1} = r_s/[r(1 - r_s/r)]$ vemos que $R^o{}_{1o1} \to \infty$ no limite $r \to r_s$ . Como as *forças de maré*[2] são proporcionais à curvatura $R^\alpha{}_{\beta\mu\nu}$ elas seriam imensamente grandes à medida que $r \to r_s$. O raio de *não−escape* $r_s = 2GM/c^2$ é denominado *raio de Schwarzschild* ou *raio gravitacional da massa* M**.** O cálculo feito usando a TGN, que dá o valor correto de $r_s$, mas nos leva a interpretar erroneamente o que ocorre: a luz ou partícula emitida radialmente para fora da região com $r \leq r_s$ não sobe, pára e depois desce. De fato, como veremos a seguir, de acordo com as previsões da TGE ela cai imediatamente e nunca começa a se movimentar radialmente para fora. A região do espaço−tempo dentro da qual um sinal pode entrar, mas da qual nenhum sinal pode sair, é chamada de *buraco negro* (BN) ou, também, *BN de Schwarzschild***.** Para um observador externo a superfície esférica de raio $r_s$ constitui−se no que chamamos de *horizonte de eventos* (HE), *horizonte de Schwarzschild* ou, simplesmente, *horizonte*. Tudo que estiver abaixo HE permanece invisível para o referido observador. Para um BN com uma massa igual à do Sol teríamos $r_s = 2GM_S/c^2 = 3.41$ km.

Levando em conta a (I.3) verifica−se que os pontos $r = r_s$ e $r = 0$ são singularidades da MS. Quando $r \to r_s$ temos $g_{oo} = (1- r_s/r) \to 0$ e $g_{11} = 1/(1- r_s/r) \to -\infty$. Quando $r \to 0$ temos $g_{oo} = (1- r_s/r) \to -\infty$ e $g_{11} = 1/(1- r_s/r) \to 0$. Conforme é visto no Apêndice B e como será mostrado na Seção 2 as singularidades dos coeficientes métricos $g_{\mu\nu}$ e da curvatura $R^{\alpha}{}_{\beta\mu\nu}$ no ponto $r_s$ podem ser eliminadas com uma escolha adequada do sistema de coordenadas onde as *forças de maré*[2] seriam finitas em $r = r_s$. Como os efeitos físicos, que são as forças de maré, permanecem finitos, bem comportados, podemos concluir que $r = r_s$ é uma singularidade matemática, espúria, ou, ainda, uma pseudo−singularidade. Ela não é uma singularidade física. Porém, o ponto $r = 0$ parece ser uma singularidade física, pois não pode ser removida por nenhuma transformação de coordenadas dentro do contexto da TGE (Apêndice B). Nas vizinhanças de $r = 0$ devem aparecer forças de maré infinitas indicando que $r = 0$ é uma singularidade física real. Talvez efeitos gravitacionais quânticos possam inibir o aparecimento dessa singularidade. [6,11]

Outro fato importante é que a singularidade no ponto $r = r_s$ gera uma diferença crítica no espaço−tempo fora e dentro do BN. Para $r > r_s$ temos $g_{oo} > 0$ e $g_{11} < 0$; para $r < r_s$, temos o contrário, $g_{oo} < 0$ e $g_{11} > 0$. Assim, se na região $r > r_s$ uma pequena mudança em t for feita mantendo com r = constante teremos $ds^2/c^2 = d\tau^2 = g_{oo}\, dt^2 > 0 \to$ separação na coordenada temporal é *timelike*. Dentro do BN, ou seja, para $r < r_s$ teremos, $ds^2/c^2 = d\tau^2 = g_{oo}\, dt^2 < 0 \to$ separação na coordenada temporal é *spacelike*. De modo análogo, se na região $r > r_s$ uma pequena mudança em r for feita mantendo t = constante teremos $ds^2/c^2 = dr^2 = g_{11}\, dr^2 > 0 \to$ separação na coordenada espacial é *timelike* e para $r < r_s$ $ds^2/c^2 = dr^2 = g_{11}\, dr^2 < 0 \to$ separação na coordenada espacial é *spacelike.*

Na Seção 1 calcularemos os tempos de percurso de um sinal luminoso e de uma sonda espacial ao descreverem uma trajetória radial no espaço−tempo descrito pela MS dada por (I.3). Na Seção 2 mostraremos como a MS (I.3) se transforma ao adotarmos as coordenadas propostas por Eddington em 1924[6,7] e por Kruskal.[10] Com essas novas coordenadas as singularidades de $g_{\mu\nu}$ e de $R^{\alpha}{}_{\beta\mu\nu}$ em $r = r_s$ desaparecem, permanecendo, entretanto, a singularidade em $r = 0$. Calcularemos os tempos de percurso da luz em percursos radiais no caso das coordenadas de Eddington e veremos como a partir da geometria de Schwarzschild surgem os buracos de minhocas (BM).

## 1. Tempos de Percursos Radiais da Luz e de Sonda na MS.

O horizonte de eventos (HE) em $r = r_s$ desempenha um papel fundamental nos BN. A região dentro desse horizonte fica rigorosamente isolada do resto do Universo. Vejamos como o HE afeta os fenômenos físicos.

O primeiro resultado importante que obtivemos[3] usando a MS foi a **dilação temporal** de um relógio com coordenada r medida por um observador muito longe do BN é dada por

$$d\tau = (1 - r/r_s)dt \qquad (1.1).$$

Assim, um relógio situado em $r \approx r_s$ anda infinitamente mais devagar do que um relógio no infinito. Isto significa que se uma sonda espacial nas proximidades de um buraco negro, os seja, com $r \approx r_s$ , enviar sinais luminosos separados por intervalos de tempo de 1 s ( medidos em seu relógio) um observador muito distante dele vai receber esses pulsos de luz separados por intervalos de tempos muito maiores do que 1s (medidos no relógio do observador).

A (1.1) dá uma dilação temporal infinita ("redshift" infinito) para um relógio em $r = r_s$. Efetivamente, como veremos a seguir, um relógio (um corpo material) não pode permanecer em repouso na superfície de eventos. Só um sinal luminoso pode permanecer em repouso em $r = r_s$.

### 1.1) *Tempo de percurso de sinal luminoso.*

Consideremos um raio de luz viajando radialmente em uma região descrita pela MS (I.3). Pondo $ds^2 = d\theta^2 = d\varphi^2 = 0$ em (I.3):

$$0 = (1 - r_s/r)c^2dt^2 - dr^2/(1 - r_s/r) \qquad (1.2).$$

de onde obtemos a **velocidade de coordenada** (ou *velocidade*) dr/dt, medida por um observador muito longe do buraco negro,

$$dr/dt = \pm\, c(1 - r_s/r) \qquad (1.3),$$

onde o sinal ± de dr/dt significa que a luz está se movimentando, respectivamente, no sentido radial positivo (SRP) ou no sentido radial negativo (SRN). Notemos que conforme (1.3), dr/dt = 0 no HE.

**a) Luz se movimentando no SRN**. Assim, usando (1.3) o tempo $t_1(r)$ que a luz leva (medido por um observador longe do BN) partindo de um ponto com coordenada inicial $r = R >> r_s$ e aproximando−se do BN até um ponto com coordenada $r \geq r_s$ é dado por

$$t_1(r) = -\int_R^r dr/c(1 - r_s/r) = (R - r)/c + (r_s/c) \ln[(R - r_s)/(r - r_s)] \quad (1.4),$$

onde $T = (R-r)/c$ seria o tempo de percurso da luz de $R \to r$ na ausência de um campo gravitacional. De acordo com (1.4) a luz levaria um tempo infinito para chegar até o ponto $r = r_s$. Para um observador muito distante a *velocidade* da luz $dr/dt \to 0$ quando $r \to r_s$ e, consequentemente, o sinal luminoso levaria um tempo infinito para chegar em $r = r_s$.

Por outro lado, o tempo de percurso $t_2(r)$ que a luz leva para ir de um ponto $r \leq r_s$ até $r = 0$ é dado por,

$$t_2(r) = -\int_r^0 dr/c(1 - r_s/r) = r/c + (r_s/c) \ln[(r_s/(r_s - r)] \quad (1.5).$$

De acordo com (1.5) o tempo de percurso $t_2$ de um sinal luminoso que parte de um ponto com $r < r_s$ é finito. Porém, ele nunca chegaria em $r = 0$ se partisse de $r = r_s$, pois nesse ponto a luz teria *velocidade* zero.

Esses resultados mostram que um sinal luminoso enviado de fora de BN em direção ao centro do BN leva um tempo infinito para chegar até o HE. Se o sinal é enviado em direção a $r = 0$ de um ponto $r < r_s$ ele levará um tempo finito para atingir o ponto $r = 0$. Ele nunca chegará ao centro se for enviado de um ponto com $r = r_s$.

**b) Luz se movimentando no SRP**. Integrando (1.5), com o sinal +, de $0 \to r_s$ vemos que o tempo de percurso de $0 \to r_s$ seria infinito, pois a *velocidade* do sinal tende a zero no HE: a luz jamais ultrapassaria a distância $r = r_s$. Porém, integrando (1.4), com o sinal +, de $r \to R$ com $r > r_s$, teríamos um tempo finito de percurso: a luz sempre chegaria até o observador que está no ponto R.

Assim, se a luz é enviada de dentro do BN no sentido $0 \to r_s$ (SRP) ela nunca ultrapassaria o HE. Porém, se ela for enviada de fora do BN de $r_s \to R$ ela sempre chegaria até um observador que está em $R > r_s$.

**1.2)** ***Tempo de Queda de uma Sonda em Percurso Radial***.

Calculemos agora o **tempo próprio** de percurso medido por um relógio colocado em uma sonda espacial que se move radialmente em direção (SRN) ao buraco negro. De acordo com os cálculos mostrados no

Apêndice C a equação (B.5) que dá a trajetória radial de uma partícula numa MS é dada por

$$(dr/cd\tau)^2 = r_s/r + 1 - B^2 \qquad (1.6),$$

onde B é uma constante de movimento. Se no estado inicial a partícula tem velocidade zero, está muito longe do BN e cai (SRN) em direção a ele, de (1.6) temos (Apêndice C)

$$d\tau = - c(r/r_s)^{1/2}\, dr \qquad (1.7).$$

Integrando (1.7) de um ponto genérico r até r = 0, temos,

$$\tau_o - \tau(r) = (2/3)(r_s/c)(r/r_s)^{3/2} \qquad (1.8).$$

onde $\tau_o$ é o tempo próprio que a sonda leva para chegar em r = 0. Notemos que o tempo próprio medido por um relógio na sonda varia suavemente ao cruzar o horizonte em $r = r_s$. Uma vez chegando no horizonte a sonda leva um tempo $\Delta\tau = (2/3)(r_s/c)$ para chegar ao centro do BN. No caso de um BN com massa 10 $M_s$ teríamos $\Delta\tau \sim 10^{-4}$ s.

Vamos agora calcular o tempo t, que a sonda levaria, medido por um observador muito longe do BN, para ir de um ponto muito distante até o horizonte de eventos. Ora, de (1.1) e (1.7) obtemos[6]

$$dt = -r^{3/2}dr/(r - r_s)r_s^{1/2} \qquad (1.9).$$

Integrando (1.9) decorre

$$t(r) = t_o + (r_s/c)[-(2/3)(r/r_s)^{3/2} - 2(r/r_s)^{1/2} + \ln|((r/r_s)^{1/2}+1)/((r/r_s)^{1/2}-1)| \qquad (1.10).$$

Para distâncias grandes teremos,

$$t(r) \approx t_o - (r_s/c)(2/3)(r/r_s)^{3/2} - 2(r_s/c)(r/r_s)^{1/2} \qquad (1.11),$$

como para grandes distâncias t(r) e τ(r) devem coincidir a constante $t_o$ deve ser escolhida adequadamente: $t_o = \tau_o + 2(r_s/c)(r/r_s)^{1/2}$ . Com essa escolha obtemos

$$t(r) \approx \tau_o - (r_s/c)(2/3)(r/r_s)^{3/2} \qquad (1.12)$$

Na Figura 1 mostramos[6] os tempos t(r) e τ(r) em função de r, representados por linhas, tracejada e contínua, respectivamente.

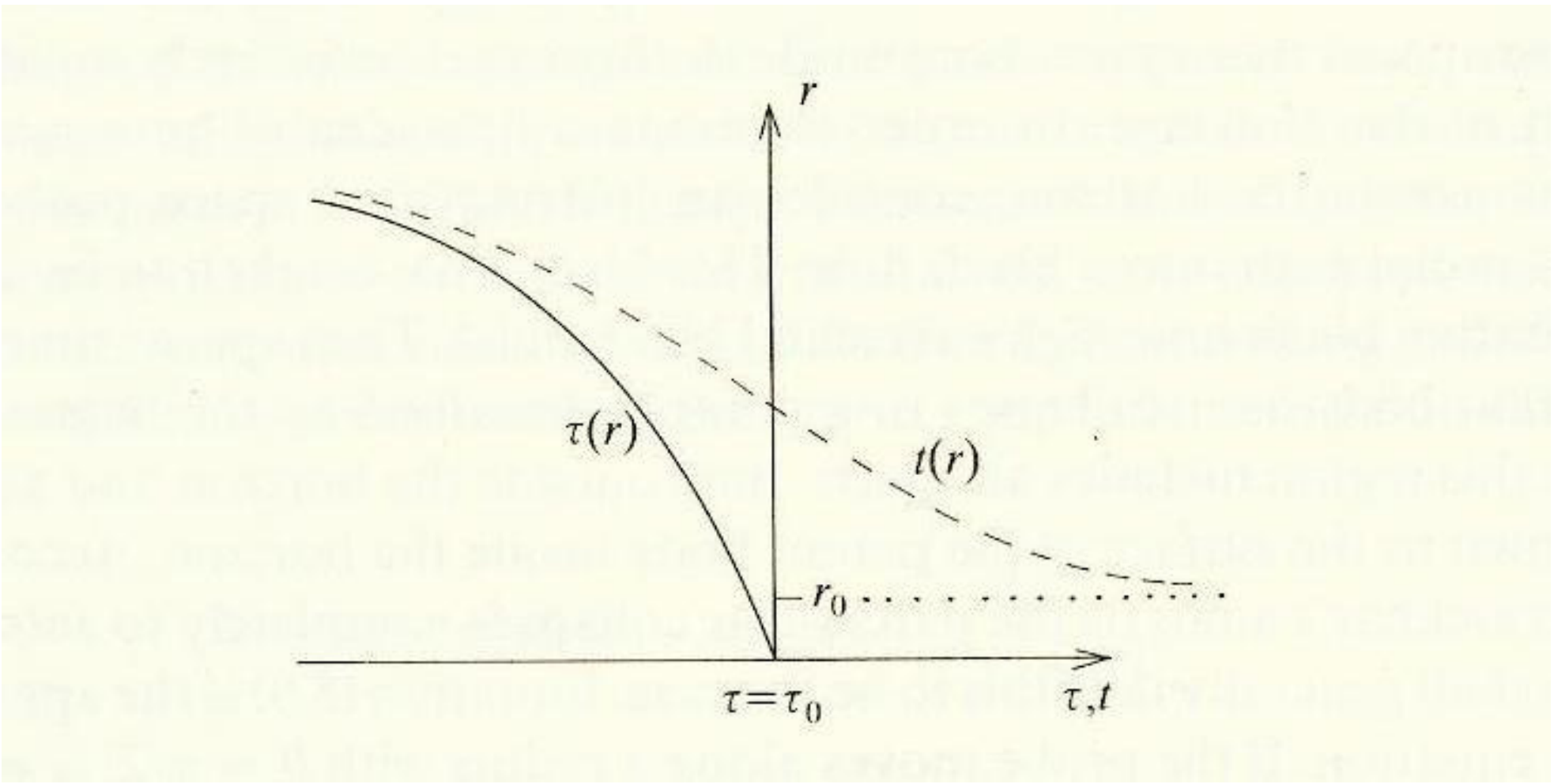

**Figura 1**. O tempo próprio τ (linha contínua) e a coordenada tempo t (linha tracejada) como uma função da coordenada radial r de uma sonda espacial caindo radialmente em um buraco negro.[6]

É importante notarmos que a sonda demora um tempo t infinito para atingir a horizonte. Ao contrário, o tempo próprio τ medido por um relógio na sonda leva um tempo finito passando pelo horizonte e atingindo r = 0. Na prática os instrumentos na sonda não sobreviveriam a essa viagem, pois seriam destruídos pelas imensas forças de maré gravitacionais.

Os comportamentos de t(r) e τ(r) em função de r são completamente diferentes quando a sonda se aproxima e atravessa o horizonte, ilustrando de modo marcante como a curvatura do espaço–tempo com a MS torna impossível cobrir todo o espaço–tempo com um único conjunto de coordenadas cartesianas.

**1.3)** ***Energia do Sinal luminoso emitido por sonda que se move com SRN.***

No item anterior (1.2) calculamos o tempo de queda de uma sonda. Suponhamos que essa sonda ao se movimentar radialmente em direção ao BN tenha uma fonte de luz que emita sinais luminosos com uma freqüência própria constante $f_o$. A energia emitida pela fonte, por unidade de tempo próprio, é a luminosidade dada $L_o \propto f_o/\Delta\tau$. Para um observador num ponto r a luminosidade $L \propto f/\Delta t$ é dada, levando em conta que $f = f_o (1 - r_s/r)^{1/2}$ e $\Delta t = \Delta\tau/(1 - r_s/r)^{1/2}$,

$$L(r) \propto (1 - r_s/r)\, f_o \qquad (1.12).$$

Considerando o tempo de chegada do sinal em um observador localizado no ponto R como sendo o instante t = 0, a luz seria emitida pela sonda no instante $-t_1(r) = T(r)$ calculado com a (1.4). Ora, de acordo com (1.4) levando em conta que a luz seja enviada em $r \approx r_s$ teremos

$$T(r) \approx -\, (r_s/c) \ln (r - r_s) \qquad (1.13).$$

De (1.12) e (1.13) obtemos,

$$L(T) \,\alpha\, f_o \exp(-\, cT/r_s) \qquad (1.14).$$

Isso mostra que o sinal emitido pela sonda decai exponencialmente á medida que ela se próxima do HE. No caso de um BN com massa 10 $M_S$ a constante de tempo $r_s/c$ é da ordem de $10^{-4}$ s. Assim, a luz branca emitida pela superfície de uma estrela que está colapsando, transformando−se em um BN, torna−se rapidamente vermelha e desaparece numa escala de tempos de $r_s/c$.

## 2. Coordenadas de Eddington e de Kruskal.

### 2.a) *Coordenadas de Eddington.*

Com o intuito de eliminar as singularidades que aparecem na MS (I.1) ou (I.3) quando se usa as coordenadas polares $x_o = x_4 = ct$, $x_1 = r$, $x_2 = \theta$ e $x_3 = \varphi$ Eddington[7] propôs em 1924 um conjunto de coordenadas mais apropriado para estudar um BN definindo um tempo t* :

$$t^* = t + (r_s/c) \ln |(r/r_s - 1)| \qquad (2.1).$$

Com essa mudança de variáveis a MS (I.3) fica substituída por

$$ds^2 = (1 - r_s/r)\, c^2 dt^{*2} - 2c(r_s/r) dr dt^* - (1 + r_s/r) dr^2 + r^2 d\Omega^2 \qquad (2.2).$$

Agora os novos coeficientes métricos $g_{\mu\nu}$ de (2.2) não têm singularidades em $r = r_s$. Apresentam ainda singularidades em $r = 0$. As coordenadas de Eddington que foram redescobertas por Finkelstein[8] em 1958 são também denominadas de coordenadas de Eddington−Finkelstein. Essas coordenadas permitem uma interpretação fisicamente mais clara do que ocorre nas vizinhanças do BM, mas, não tão óbvias para regiões mais distantes do BN. Pode−se mostrar (Apêndice B) que com as novas coordenadas não há singularidades na curvatura $R^{\alpha}{}_{\beta\mu\nu}$ em $r = r_s$ e, consequentemente, as forças de maré são finitas nesse ponto. Entretanto (Apêndice B) as forças de maré tendem a infinito quando $r \to 0$.

Nessas novas coordenadas a trajetória radial de um sinal luminoso é determinada a partir de

$$0 = (1 - r_s/r)\, c^2 dt^{*2} - 2c(r_s/r) dr dt^* - (1 + r_s/r) dr^2$$

que tem como soluções

$$dr/dt^* = -c \qquad e \qquad dr/dt^* = c\,(1-r_s/r)/(1+r_s/r) \quad (2.3).$$

Na ausência de um BN as (2.3) ficam dadas por $dr/dt^* = -c$ e $dr/dt^* = c$, respectivamente. A primeira equação descreveria um sinal luminoso indo em direção a $r = 0$ (SRN) e a segunda um sinal indo em direção contrária a $r = 0$ (SRP).

Assim, no caso geral, a primeira equação de (2.3) descreveria um sinal luminoso que se move em direção a $r = 0$ com uma *velocidade* constante $-c$ para qualquer valor de r, ou seja, $0 \leq r < \infty$. Em outras palavras a luz que vem de fora sempre vai entrar no BN. A segunda solução que descreveria a trajetória da luz no SRP mostra que para $0 \leq r < r_s$ a *velocidade* é negativa, para $r = r_s$ a *velocidade* é nula e que a partir de $r > r_s$ ela se torna positiva e vai crescendo até atingir o valor máximo $+ c$ no infinito. Assim, se a sonda estiver dentro do horizonte a luz que ela emite terá sempre *velocidade* voltada para $r = 0$ ou SRN. Desse modo a luz que vem de fora (SRN) ou a que é emitida de dentro do BN no SRP ou SRN sempre vai, inexoravelmente, em direção a $r = 0$.

Há uma forma alternativa de coordenadas de Eddington que é, ao invés de (2.1), definir

$$t^* = t - (r_s/c) \ln |(r/r_s - 1)| \qquad (2.4).$$

Com essa escolha as (2.3) são substituídas por

$$dr/dt^* = c \qquad e \qquad dr/dt^* = -c\,(1-r_s/r)/(1+r_s/r) \qquad (2.5).$$

Nessas condições pode−se verificar que a luz sempre sairá do horizonte e ao invés de um BN temos um *buraco branco* (BB**).** Desse modo a matéria sempre será ejetada da singularidade $r = 0$. Conforme teorias vigentes de evolução estelar[2,6,9] somente BN devem existir e nunca BB, embora matematicamente isso fosse possível de acordo com (2.5). Isso mostra que escolhas de coordenadas num espaço−tempo descrito pela MS podem gerar propriedades muito sutis como veremos a seguir analisando as coordenadas de Kruskal.[10]

## 2.b) *Coordenadas de Kruskal.*

Um outro sistema de coordenadas $(u,v,\theta,\varphi)$ muito conveniente para estudar os BN foi proposto por Kruskal[10]. Para o interior (a) e para o exterior (b) do BN temos,

(a) $r < r_s$

$$u = (1 - r/r_s)^{1/2} \exp(r/2r_s) \sinh(ct/2r_s)$$
$$v = (1 - r/r_s)^{1/2} \exp(r/2r_s) \cosh(ct/2r_s) \qquad (2.6)$$

(b) $r > r_s$

$$u = ( r/r_s - 1)^{1/2} \exp(r/2r_s) \cosh(ct/2r_s)$$
$$v = ( r/r_s - 1)^{1/2} \exp(r/2r_s) \sinh(ct/2r_s) \qquad (2.7).$$

As transformações inversas de (a) são dadas por

$$( r/r_s - 1) \exp(r/r_s) = u^2 - v^2 \quad \text{e} \quad ct = 2r_s \tanh^{-1}(u/v) \qquad (2.8),$$

e as inversas de (b) dadas por

$$( r/r_s - 1) \exp(r/r_s) = u^2 - v^2 \quad \text{e} \quad ct = 2r_s \tanh^{-1}(v/u) \qquad (2.9).$$

Com essas novas coordenadas de Kruskal o intervalo $ds^2$ definido por (I.3) fica escrito como

$$ds^2 = (4r_s{}^3/r) \exp(-r/r_s) (dv^2 - du^2) - r^2 d\theta^2 - r^2 \sin^2\theta \, d\varphi^2 \qquad (2.10),$$

onde a coordenada r deve ser vista como uma função de u e v , conforme (2.8) ou (2.9). Como no caso da métrica de Eddington, com métrica de Kruskal (2.10) só restou a inevitável singularidade $r = 0$. Esta última é também chamada de "extensão *maximal* da métrica de Schwarzschild". A variedade definida pela métrica de Kruskal é *maximal,* pois nela as geodésicas têm um comprimento infinito em ambas as direções (não têm começo nem fim) ou começam, ou terminam, em uma singularidade.

Nos livros do Misner, Thorne and Wheeler[10] vemos em detalhes como obter as geodésicas radiais das partículas massivas e dos fótons usando o plano (u,v) de Kruskal comparando com as obtidas usando as coordenadas de Schwarzschild (ct,r) (vide Ohanian[2]). As geodésicas radiais de fótons ("lightlike") são obtidas de (2.10) fazendo $ds^2 = 0$, $\theta = \pi/2$ e $\varphi$ = constante, obtendo $dv^2 = du^2$. Mostrando que as geodésicas dos fótons são retas $u = v$ no plano (u,v). Diferenciando (2.6) e (2.7) verifica−se que ao longo das geodésicas temos $dr/dt = \pm c$. Na Figura 2 mostramos uma das representações do plano (u,v) que aparece no livro do Ohanian[2] [Fig.(9.4), pag.319]. No plano (u,v) aparecem as regiões I e IV do espaço−tempo (u,v) que representam, respectivamente, o BN e o BB, que tem o ponto $u = v = 0$ em comum. As regiões I e III estão fora do cone de luz: é impossível haver comunicação entre elas.

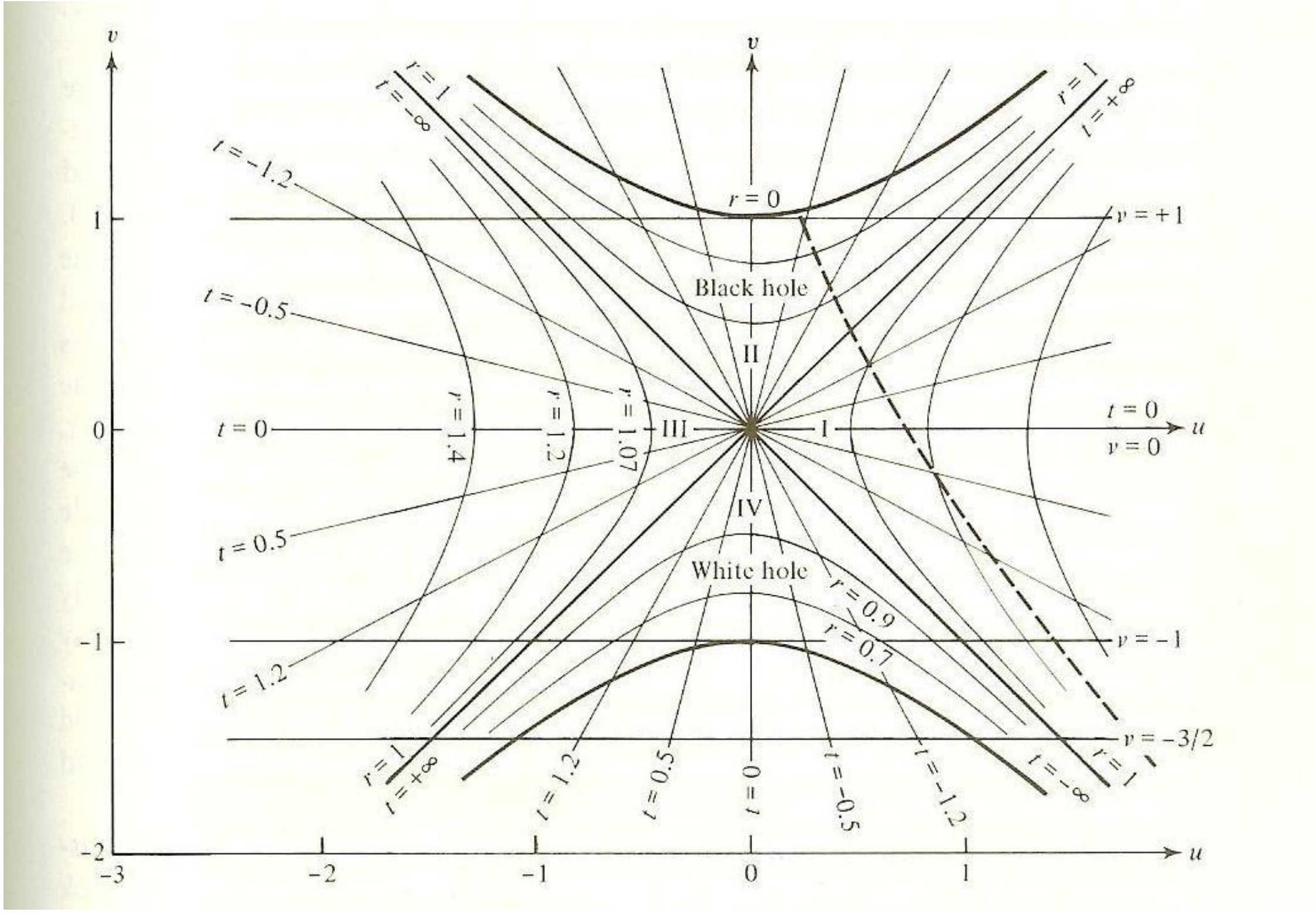

**Figura 2**. A geometria[6] maximal de Schwarzschild nas coordenadas (u,v) de Kruskal.

2.c) *Aspectos da Geometria do Espaço−Tempo de Schwarzschild: o Buraco de Minhoca.*

É fascinante analisarmos a geometria do espaço−tempo envolvida com os fenômenos físicos. Sugerimos aos alunos que leiam a seção 23.8 do livro Gravitation[11] e os artigos de Fuller, Misner e Wheeler[16,17] sobre esse tema, a "*geometrodinâmica*". Para estrelas estáticas muito relativísticas a geometria do espaço−tempo se desvia fortemente da geometria plana Lorentz−Euclideana. Nessas condições verifica−se, levando em conta (I.3), com t = constante, que a distância radial $\ell(r)$ de um ponto de coordenada r medida a partir de $r = r_s$ é dada por

$$\ell(r) = \int_{r_s}^{r} dr/(1 - r_s/r)^{1/2},$$

a área A(r) de uma esfera com raio r é dada por $A(r) = 4\pi r^2$ e o comprimento s(r) da circunferência de raio r medida no plano equatorial onde $\theta = \pi/2$ é dado por $s(r) = 2\pi r$. Como $(1 - r_s/r)^{1/2} < 1$ verificamos que $d\ell(r)/dr$ varia muito rapidamente para r nas proximidades de $r_s$ e fica constante no limite de $r >> r_s$. Esse comportamento "estranho" pode ser visualizado mais facilmente usando um processo denominado de "imersão geométrica".

Fazendo $\theta = \pi/2$ em (I.3) teremos um elemento de arco $dl^2$ no plano equatorial dado por

$$dl^2 = dr^2/(1- r_s/r) + r^2 d\varphi^2 \qquad (2.11),$$

que obedece a uma geometria de um espaço 2−dim curvo $(r,\varphi)$. O que se faz é emergir esse espaço curvo em um espaço 3−dim plano $(r,\varphi,z)$. Nesse espaço a coordenada z é uma "coordenada artificial" que não tem nada a ver com a coordenada z do espaço real. As distâncias dl no espaço 3−dim plano com coordenadas cilíndricas $(r,\varphi,z)$ onde iremos emergir a curva 2−dim descrita por (2.11) são dadas por

$$dl^2 = dr^2 + r^2 d\varphi^2 + dz^2 = dr^2 [1 + (dz/dr)^2] + r^2 d\varphi^2 \qquad (2.12).$$

A imersão é feita de tal modo que as distâncias dl ao longo da superfície descrita por (2.12) coincidam com as distâncias dl dadas por (2.11). Com essa condição, de (2.11) e (2.12), obtemos

$$(dz/dr)^2 = 1/(1- r_s/r) - 1 \qquad (2.13).$$

Integrando (2.13) teremos para $z > 0$

$$z(r) = \int dr/(r/r_s - 1)^{1/2} = 2r_s(r/r_s - 1)^{1/2} + \text{constante} \qquad (2.14).$$

Supondo que em $z = 0$ tenhamos $r = r_s$ obtemos a superfície de um parabolóide de revolução dada por

$$z(r) = 2r_s (r/r_s - 1)^{1/2} \qquad (2.15).$$

A parte de baixo da superfície, ou seja, a parte com $z < 0$, pode ser pensada como uma deformação semelhante de um "segundo universo".[2,11,16,17] Na Figura 3 (vide fig.9.5(a) do Ohanian[2], pág.322) vemos a superfície parabólica z(r) conectando os "dois universos"(dois espaços planos) Euclideanos . Ela se alarga, abrindo−se nas duas extremidades como um funil e se estreita no meio em $z = 0$. Essa superfície denomina−se "buraco de minhoca (BM)" (ou "wormhole").[2,11,16,17] As partes de cima e de baixo do funil são superfícies planas que representam o espaço plano Euclideano muito distante do BN que estaria no centro ($z = 0$), no gargalo do funil. Nesse ponto mais estreito $r = r_s$, nas proximidades do BN, ocorre a maior deformação do espaço−tempo.

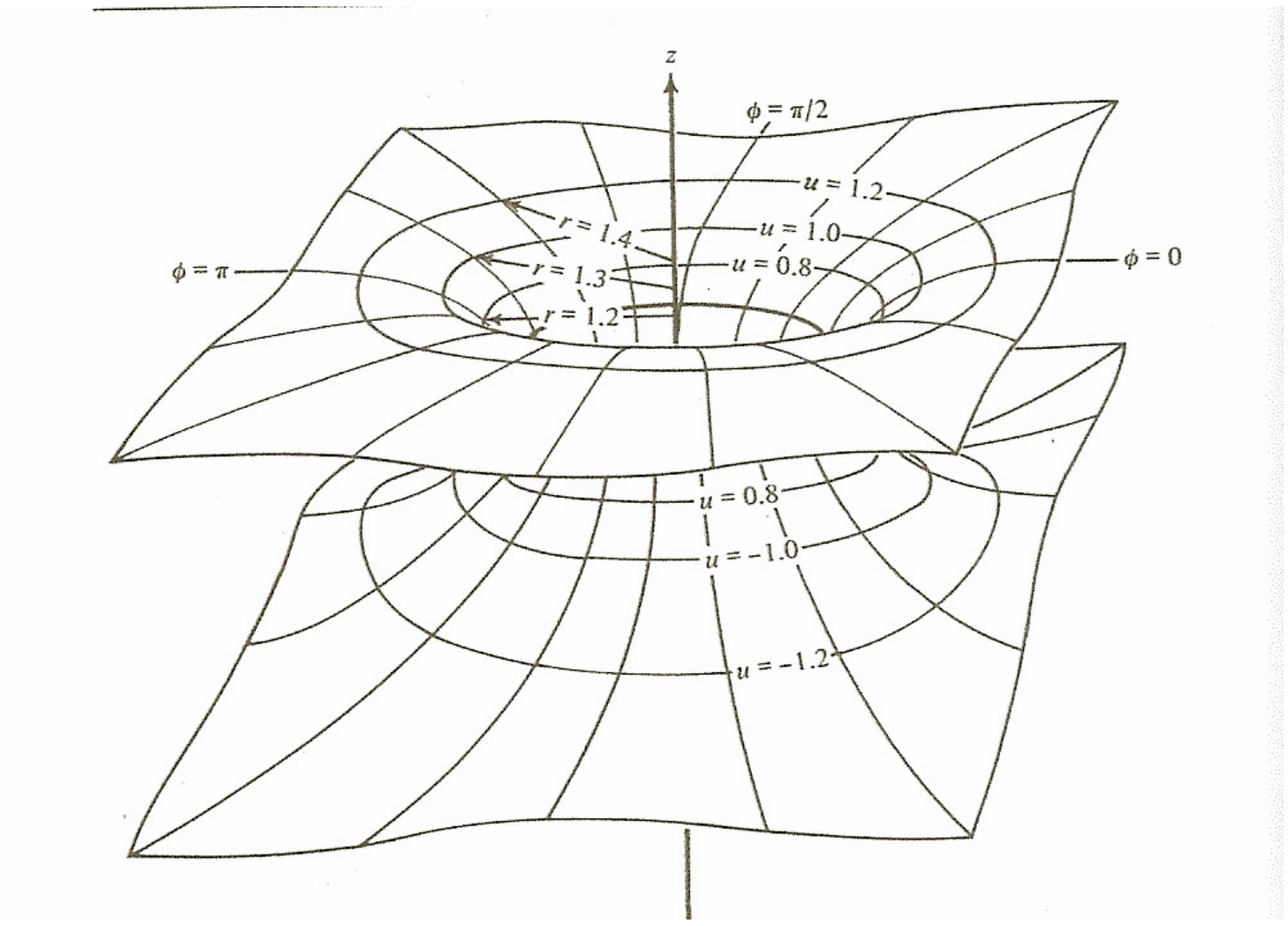

**Figura 3**. Geometria de um Buraco de Minhoca (BM).[2] A superfície $z(r,\varphi)$ foi calculada para o caso $t = 0$ e $\theta = \pi/2$ (na métrica de Kruskal temos $v = 0$ e $\theta = \pi/2$). A coordenada r é medida em unidades de $r_s$ e o parâmetro u é dado, pondo $t = 0$ em (2.7), por $u = \pm(r/r_s - 1)^{1/2} \exp(r/2r_s)$.

Sobre a superfície 2−dim z(r) as distâncias medidas $d\ell$ entre quaisquer dois pontos próximos $(r,\varphi)$ e $(r + dr, \varphi + d\varphi)$ estão corretamente reproduzidas. Os círculos de raios r tem circunferências próprias s(r) iguais a $2\pi r$. Distâncias medidas fora da superfície não tem significado físico; pontos fora da superfície não têm nenhum significado físico; o espaço 3−dim Euclideano não tem nenhum significado físico. Somente a superfície curva 2−dim tem significado. As regiões 3−dim dentro e fora do funil não têm significado físico, ou seja, o espaço "imersor" Euclideano $(r,\varphi,z)$ não tem significado físico. Ele somente permite visualizar a geometria do espaço em torno da estrela de um modo conveniente: com ela podemos visualizar quão rapidamente as distâncias $\ell$ crescem em função das coordenadas $(r,\varphi)$ e como as circunferências (secções retas do funil) s(r) variam com r.

Interpreta−se também o BM como sendo uma conexão entre um único espaço plano Euclideano[10,11,16] no caso limite em que as bocas dos funis estão muito distantes umas das outras comparativamente às dimensões dos gargalos do BM.

A métrica de Kruskal depende do tempo t que aparece nas funções u e v conforme vemos em (2.6)−(2.9). Na Figura 3 mostramos caso do BM (BM) para o instante $t = 0$ ou $v = 0$ ($v > 0$ quando $t > 0$ e $v < 0$ quando $t < 0$). A coordenada v faz o papel do "tempo" na métrica de Kruskal.[2] Variando t consequentemente u e v variam. Pode−se mostrar[17] que a geometria do BM varia com v (ou t), conforme esquematizamos[2] na Fig.4. Para v < −1 não existe um BM, somente dois espaços Euclideanos inferior e superior,desconectados, cada um com uma cúspide: o BM está colapsado.

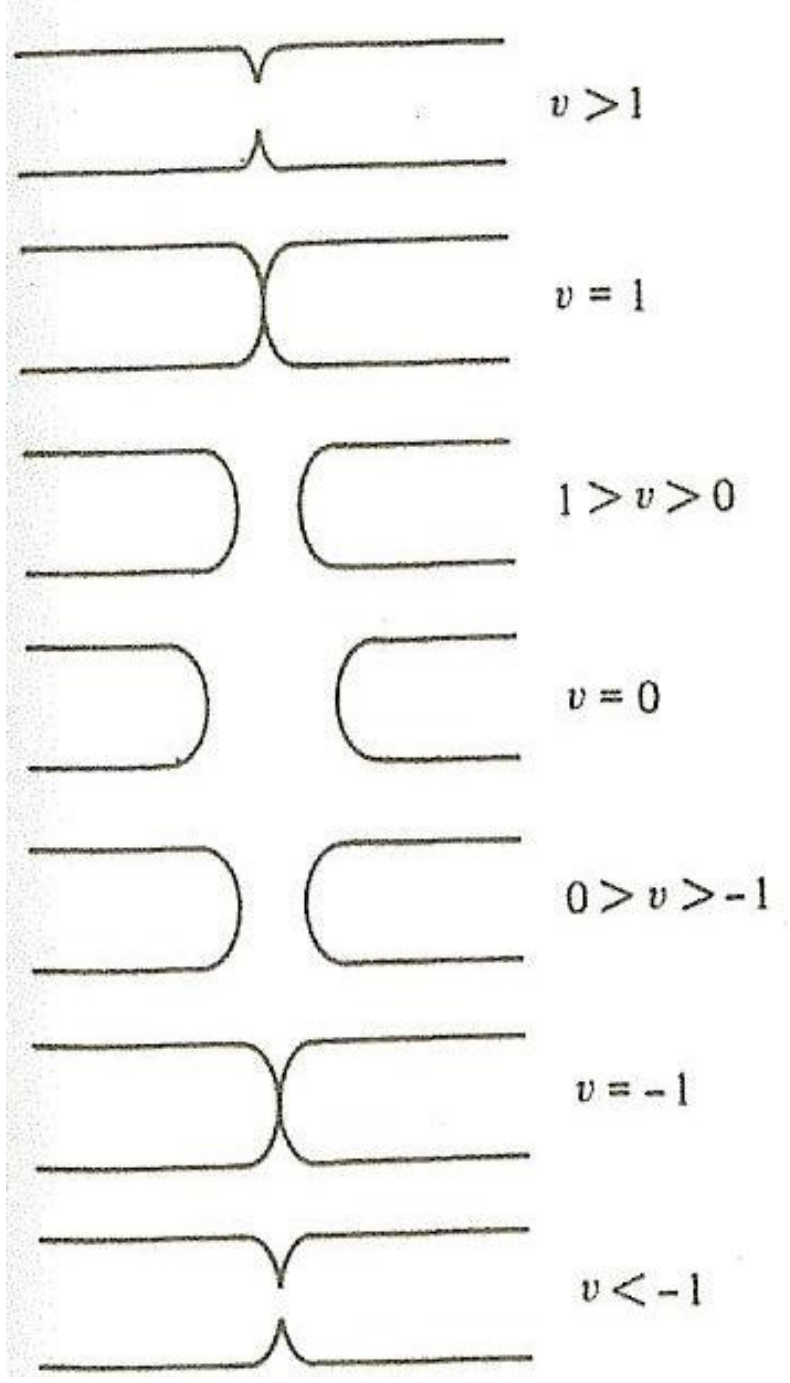

**Figura 4.** Esquema[2] da evolução temporal da geometria do Buraco de Minhoca (BM).

Para v = −1 o BM passa a existir, mas, o gargalo está fechado. Para 0 > v > −1 o gargalo está aberto e para v = 0, que é o caso visto na Fig.3, ele tem um diâmetro máximo. Para tempos positivos 1 > v > 0 o gargalo começa a se fechar, fechando−se em v = 1. Para v > 1 o BM colapsa deixando desconectados os espaços Euclideanos inferior e superior, cada um com uma cúspide.

Finalmente, é importante recordarmos que as métricas de Schwarzschild e de Kruskal são válidas numa região onde não há matéria, ou seja, são soluções das equações da TGE numa região onde $T_{\nu\mu} = 0$. Assim, não são relevantes para o problema do colapso gravitacional que deve dar origem aos BN. As interpretações deduzidas usando essas métricas, inclusive os BM, podem ser aplicadas somente para o caso dos BN já existentes no Universo.

## 3. Comentários.

O Universo observado até atualmente parece indicar que talvez seja muito mais provável que não existam BN de Schwarzschild e que se eles existirem deverão ser muito mais parecidos com os BN de Kerr[2]. Porém, não será nenhuma surpresa se os BN reais tiverem muito pouca semelhança com os dois tipos citados acima.

**APÊNDICE A.** *Órbita Circular da Luz em torno de uma Estrela muito Massiva e Compacta.*

Conforme artigos anteriores[1,3] para calcular a trajetória (geodésica) de uma partícula que se move no vácuo temos de levar em conta as seguintes equações

$$ds^2 = g_{\mu\nu}\, dx^\mu\, dx^\nu$$

e (A.1)

$$d^2x^\alpha/ds^2 \; + \; \Gamma_{\tau\nu}{}^\alpha\, (dx^\nu/ds)\, (dx^\tau/ds) \; = 0,$$

onde os símbolos $\Gamma_{\mu\nu}{}^\alpha = \{_\mu{}^\alpha{}_\nu\}$ de Christoffel são definidos por

$$\Gamma_{\mu\nu}{}^\alpha = \{_\mu{}^\alpha{}_\nu\} = (g^{\alpha\lambda}/2)(\, \partial_\nu\, g_{\lambda\mu} + \partial_\mu\, g_{\lambda\nu} - \partial_\lambda\, g_{\mu\nu}\,) \qquad \text{(A.2)}$$

e os tensores $g_{\mu\nu}$ são definidos através da métrica de Schwarzschild

$$ds^2 = e^{N(r)}\, c^2dt^2 \; - (e^{M(r)}dr^2 + r^2d\theta^2 + r^2\sin^2\theta\; d\varphi^2) \qquad \text{(A.3)},$$

onde $e^{N(r)} = g_{oo}(r) = g_{44}(r) = Z = (1 - 2\kappa/r)$, $e^{M(r)} = g_{11}(r) = 1/Z$, $g_{22}(r) = r^2$ e $g_{33}(4) = r^2\sin^2\theta$ , levando em conta que $x_o = x_4 = ct$, $x_1 = r$, $x_2 = \theta$ e $x_3 = \varphi$. Com esses valores podemos calcular os símbolos de Christoffel. Lembrando[4] que $g^{\mu\nu} = M_{\mu\nu}/|g|$ onde g é o determinante de $g_{\mu\nu}$ e $M_{\mu\nu}$ é o determinante menor de $g_{\mu\nu}$ em g. Como os elementos de g são diagonais temos $|g| = |g_{11}\, g_{22}\, g_{33}\, g_{44}| = c^2e^{2(N+M)}\, r^4\sin^2\theta$. Assim, $g^{oo} = c^{-2}e^{-N}$, $g^{11} = e^{-M}$ , $g^{22} = r^{-2}$ e $g^{33} = r^{-2}\sin^{-2}\theta$. Como os $g_{\mu\nu}$ só dependem de $r = x_1$ em (A.2) só há derivadas do tipo $\partial_r\, g_{\mu\nu} = \partial_1 g_{\mu\nu}$. Indicando por $N' = \partial N/\partial\, r$ e $M' = \partial M/\partial\, r$ obtemos $\Gamma_{\mu\nu}{}^\alpha = \{_\mu{}^\alpha{}_\nu\}$, seguindo o mesmo procedimento visto em detalhes num artigo anterior.[1,3,12,13]

No caso particular da trajetória de um sinal luminoso temos uma "geodésica nula"[13,14], ou seja, devemos assumir que $ds^2 = 0$. Neste caso defini−se um parâmetro escalar $\lambda$ não nulo que varia ao longo dessa geodésica. Assim, as equações mostradas em (A.1) e (A.3) são substituídas, respectivamente, por

$$g_{\mu\nu}(dx^\mu/d\lambda)(dx^\nu/d\lambda) = 0 \qquad \text{(A.4)}$$

$$d^2x^\alpha/d\lambda^2 \; + \; \Gamma_{\tau\nu}{}^\alpha\, (dx^\nu/d\lambda)\, (dx^\tau/d\lambda) \; = 0, \qquad \text{(A.5)}$$

$$0 \; = Z\, c^2\, (dt/d\lambda)^2 \; - [Z^{-1}\, (dr/d\lambda)^2 + r^2\, (d\theta/d\lambda)^2 + r^2\sin^2\theta\, (d\varphi/d\lambda)^2]. \qquad \text{(A.6)}$$

Usando a (A.5) e os $\Gamma_{\mu\nu}{}^\alpha$ calculados conforme dissemos acima obtemos as seguintes equações [12]

$$d\{Z(dt/d\lambda)\}/d\lambda = 0, \quad (A.7)$$

$$d\{r^2 (d\theta/d\lambda)\}/d\lambda - r^2 \sin\theta \cos\theta \, (d\varphi/ d\lambda)^2 = 0 \quad (A.8)$$

$$d\{r^2 \sin^2\theta \, (d\varphi/ d\lambda)\}/d\lambda = 0 \quad (A.9).$$

Assumindo que o movimento da luz se efetue num plano,[3] pomos $d\theta/d\lambda = 0$ e $\theta = \pi/2$ em (A.7) e (A.9), obtendo delas, respectivamente,

$$dt/d\lambda = \beta^*/Z \qquad e \qquad r^2 (d\varphi/d\lambda) = h\,\beta^* \quad (A.10),$$

onde h e $\beta^*$ são constantes de integração.[3] Substituindo (A.10) em (A.6) teremos, fazendo $u = 1/r$,

$$(du/d\varphi)^2 = 1/h^2 - u^2 + 2\kappa u^3 \quad (A.11).$$

Derivando (A.11) em relação a $\varphi$ deduzimos a seguinte equação,[2,6]

$$d^2u/d\varphi^2 + u = 3\kappa u^2 \quad (A.12).$$

Esta equação admite como solução $r$ = constante = $3\kappa = 3GM/c^2$. Isto implica que a luz pode assumir uma órbita circular com raio $3GM/c^2$ em torno da estrela. Se u for um pouco maior (menor) do que $(3GM/c^2)^{-1}$ a órbita é instável.

## APÊNDICE B. *Comentários sobre as Singularidades na Métrica de Schwarzschild e as Forças de Maré.*

A *curvatura* $R^\alpha{}_{\beta\mu\nu}$ do espaço–tempo de Riemann ou *tensor de Riemann–Christoffel*, é definida por[1,2,6,12,13]

$$R^\sigma{}_{\lambda\mu\nu} = \partial^\mu \Gamma_{\lambda\nu}{}^\sigma - \partial^\nu \Gamma_{\lambda\mu}{}^\sigma + \Gamma_{\lambda\nu}{}^\tau \Gamma_{\mu\tau}{}^\sigma - \Gamma_{\lambda\mu}{}^\tau \Gamma_{\tau\nu}{}^\sigma \quad (B.1),$$

onde $\Gamma_{\mu\nu}{}^\alpha = \{_\mu{}^\alpha{}_\nu\}$ são os *símbolos de Christoffel* definidos por

$$\Gamma_{\mu\nu}{}^\alpha = \{_\mu{}^\alpha{}_\nu\} = (g^{\alpha\lambda}/2)(\partial_\nu g_{\lambda\mu} + \partial_\mu g_{\lambda\nu} - \partial_\lambda g_{\mu\nu}) \quad (B.2).$$

As equações de campo da TGE são dada por[2,6,12,13],

$$R_{\mu\nu} - (1/2) g_{\mu\nu} R = \kappa\, T_{\mu\nu}{}^{(m)} \quad (B.3),$$

onde $\kappa = 8\pi G/c^4$, $R_{\mu\nu}$ é o *tensor de curvatura de Ricci* definido por

$$R_{\mu\nu} = R_{\nu\mu} = \partial_\mu \Gamma_{\nu\sigma}{}^{\sigma} - \partial_\sigma \Gamma_{\nu\mu}{}^{\sigma} + \Gamma_{\mu\sigma}{}^{\tau} \Gamma_{\nu\mu}{}^{\sigma} - \Gamma_{\nu\mu}{}^{\tau} \Gamma_{\tau\sigma}{}^{\sigma} \qquad \text{(B.4)},$$

e o escalar $R = g^{\lambda\sigma} R_{\sigma\lambda}$ é conhecido como *curvatura escalar* ou *invariante de curvatura* do espaço–tempo. Pode –se mostrar[2] que as *forças de maré***,** que são efeitos físicos, são diretamente proporcionais a $R^{\sigma}{}_{\lambda\mu\nu}$. Suponhamos que em num dado sistema de coordenadas existam pontos singulares nos quais $R^{\sigma}{}_{\lambda\mu\nu} \rightarrow \infty$. Se existir uma adequada transformação de coordenadas que permita eliminar essas singularidades, ou seja, de tal modo que os efeitos físicos, que são as forças de maré, permanecem finitos, bem comportados, podemos concluir que as singularidades são espúrias, matemáticas ou, ainda, pseudo–singularidades.

Por exemplo, verifica–se[2] que na MS (I.3) a componente $R^{o}{}_{101}$ é dada por $R^{o}{}_{1o1} = (r_s/r)/(1 - r_s/r)$ que tende a infinito no limite $r \rightarrow r_s$. Usando, por exemplo, as coordenadas de Eddington–Filkenstein (vide Cap.2) e as (B.1) – (B.4) pode–se mostrar que as forças de maré são finitas no ponto $r = r_s$. Somente no ponto $r = 0$ elas divergem. Um outro sistema de coordenadas para analisar os BN seria as "*coordenadas geodésicas*" (vide Ohanian,[2] pág.309) que são as que apresentam num certo ponto P uma métrica de espaço–plano. É sempre possível[1,2] encontrar um sistema de coordenadas que obedeçam a essa condição. Isto pode ser visto, por exemplo, no livro do Ohanian[2] (pág. 231– 233). Em outras palavras dadas as coordenadas $x^\mu$ com uma métrica $g_{\mu\nu}(x)$ podemos sempre encontrar uma transformação linear para novas coordenadas $x'^\mu = b^\mu{}_\nu x^\nu$, onde $b^\mu{}_\nu$ são constantes, de tal modo que num certo ponto P tenhamos $g'_{\mu\nu}(P) = \eta_{\mu\nu}$, onde $\eta_{\mu\nu}$, = (–1,1,1,1) característico de um espaço–plano de Minkowski. Nas proximidades do ponto P o espaço é localmente plano onde temos um *referencial inercial local*. Para uma partícula no ponto P temos $d^2x'^\mu/d\tau = 0$, ou seja, ela se move com velocidade constante ou permanece em repouso vista no sistema de coordenadas geodésicas, o qual estaria, então, instantaneamente em *queda livre* com a mesma aceleração da partícula. O ponto P é tomado como origem do *referencial geodésico* ou em *queda livre*. Enfatizamos que as coordenadas são geodésicas *somente* para um determinado instante. As derivadas de $g'_{\mu\nu}(P)$ são zero somente num ponto P do espaço–tempo. Isto é, em *um lugar* e em *um tempo*. Se quisermos obter coordenadas geodésicas em outro ponto P´ torna–se necessário efetuar uma nova transformação de coordenadas para esse outro ponto P´ do espaço–tempo.

O espaço–tempo fica plano numa vizinhança infinitesimal do ponto P na qual as derivadas primeiras ordem de $g'_{\mu\nu}(x')$ se anulam. Por outro lado, as derivadas segundas de $g'_{\mu\nu}(x')$ não podem ser *todas* anuladas por nenhuma transformação de coordenadas.[2] Isto implica que as forças de maré não se anulam nas *vizinhanças* de P. Essas forças poderiam ser medidas o que permitiria discriminar entre o efeito criado por um campo gravitacional do efeito gerado por uma pseudo força de um campo de aceleração. Nesse aspecto os efeitos de um campo de gravitação não são indistinguíveis dos efeitos observados em um referencial acelerado. Isso poderia ocorrer somente no caso em que o campo gravitacional é *perfeitamente uniforme* quando a força de maré seria exatamente igual a zero. Vide comentários sobre isso (que envolve o *Princípio de Equivalência*), por exemplo, no livro do Ohanian[2] nas págs.38–41.

Para podermos afirmar que um ponto é uma singularidade física precisamos verificar quantitativamente os seus valores em diferentes coordenadas. Entretanto, de acordo com o *invariante de Kretschmann*[18] que para um BN de Schwarschild é dado por

$$R^{\alpha\beta\gamma\delta} R_{\alpha\beta\gamma\delta} = 12\, r_s^2/r^6 \qquad \text{(B.5)},$$

a singularidade em r = 0 deve existir sempre, independentemente do sistema de coordenadas escolhido. Talvez efeitos gravitacionais quânticos possam inibir o aparecimento dessa singularidade.[6,11]

Em vários aspectos a singularidade $r = r_s$ que aparece na MS (I.3) é similar[2,13] a que encontramos em um sistema de coordenadas em rotação no caso de um espaço–tempo plano. Consideremos um espaço plano de Lorentz definido pelo elemento de linha $ds^2$ escrito, respectivamente, em coordenadas cartesianas ($x_o = x_4 = ct$, $x_1 = x$, $x_2 = y$ e $x_3 = z$) ou em polares cilíndricas ($x_o = x_4 = ct$, $x_1 = r$, $x_2 = \varphi$ e $x_3 = z$) :

$$ds^2 = c^2\, dt^2 - dx^2 - dy^2 - dz^2 = c^2\, dt^2 - dr^2 - r^2\, d\varphi^2 - dz^2 \qquad \text{(B.6)}.$$

No caso cartesiano $g_{oo} = g_{44} = 1$ e $g_{11} = g_{22} = g_{33} = -1$ e no caso polar $g_{oo} = g_{44} = 1$, $g_{11} = -1$, $g_{22}(r) = -r^2$ e $g_{33} = -1$. Assim, em coordenadas cartesianas temos $R^{\sigma}{}_{\lambda\mu\nu} = 0$ e , consequentemente, também em coordenadas polares o tensor é nulo pois o referencial polar é obtido a partir do cartesiano através de uma transformação definida por ct = ct, x = r cosφ , y = r sinφ e z = z.

Passando de um referencial inercial para um não inercial em rotação com velocidade angular constante ω em torno do eixo z obtemos de (B.6) com uma transformação de coordenadas $x_4 = ct$, $x_1 = r$, $x_2 = \varphi + \omega t$ e $x_3 = z$ :

$$ds^2 = (1 - \omega^2 r^2/c^2) c^2\, dt^2 - dr^2 - r^2\, d\varphi^2 - dz^2 - 2\omega r^2\, d\varphi\, dt \qquad \text{(B.7)},$$

onde $g_{oo} = (1 - \omega^2 r^2/c^2)$, $g_{11} = g_{33} = -1$, $g_{13} = g_{31} = -r^2$ e $g_{02} = g_{20} = -2\omega r^2$. Obviamente $g_{oo}$ é singular no ponto r = c/ω que define uma superfície com "redshift" infinito. Entretanto, o tensor de curvatura $R^{\sigma}{}_{\lambda\mu\nu} = 0$ também no caso de um referencial em rotação, pois, (B.7) foi obtida por uma transformação de coordenadas a partir de (B.6).

## APÊNDICE C. *Trajetória de uma Partícula Massiva em um Campo Gravitacional.*

Levando em conta que na MS (I.1) temos $g_{oo}(r) = g_{44}(r) = Z = (1-2\kappa/r)$, $g_{11}(r) = 1/Z$, $g_{22}(r) = r^2$ e $g_{33}(4) = r^2\sin^2\theta$ o invariante[14,15]

$$g_{\mu\nu}(dx^{\mu}/ds)(dx^{\nu}/ds) = 1 \qquad (C.1)$$

que está associado à equação da trajetória (geodésica) de uma partícula com massa $\neq 0$ fica escrito como,

$$Z\,(dt/cd\tau)^2 - (dr/cd\tau)^2/Z - r^2\,(d\theta/cd\tau)^2 - r^2\sin^2\theta(d\varphi/cd\tau)^2 = 1 \qquad (C.2).$$

Efetuando cálculos análogos aos que utilizamos para obter a trajetória de um planeta em torno do Sol podemos mostrar,[3] para um movimento que se efetua em um plano com $\theta = \pi/2$ = constante, que

$$r^2 d\varphi/d\tau = A \qquad \text{e} \qquad dt/d\tau = B/Z \qquad (C.3),$$

onde A = velocidade areolar e B são constantes de integração. Nessas condições a (C.2) fica escrita como

$$B^2/Z^2 - (dr/cd\tau)^2/Z^2 - A^2/r^2 = 1 \qquad (C.4).$$

No caso em que partícula se move radialmente A = 0 a (C.4) dá

$$(dr/cd\tau)^2 = (r_s/r) - 1 + B^2 \qquad (C.5).$$

No caso em que a partícula está inicialmente muito distante, $r >> r_s$ , e em repouso, através de (c.5), verificamos que B = 1.

**REFERÊNCIAS**